# Confocal laser scanning microscopy image correlation for nanoparticle flow velocimetry

Brian Jun[1], Matthew Giarra[2], Haisheng Yang[3], Russell Main[3] and Pavlos Vlachos[1]

[1] *Department of Mechanical Engineering, Purdue University, West Lafayette, IN, USA*

[2] *Department of Mechanical Engineering, Virginia Tech, Blacksburg, VA, USA*

[3] *Department of Basic Medical Sciences, Purdue University, West Lafayette, IN, USA*

Communication author:

Pavlos P Vlachos

Email: pvlachos@purdue.edu

Word count : 6650

Number of figures: 13

Number of tables: 2

**Abstract**

We present a new particle image correlation technique for resolving nanoparticle flow velocity using confocal laser scanning microscopy (CLSM). The two primary issues that complicate nanoparticle scanning laser image correlation (SLIC)–based velocimetry are (1) the use of diffusion-dominated nanoparticles as flow tracers, which introduce a random decorrelating error into the velocity estimate, and (2) the effects of the scanning laser image acquisition, which introduces a bias error. To date, no study has quantified these errors or demonstrated a means to deal with them in SLIC velocimetry. In this work, we build upon the robust phase correlation (RPC) and existing methods of SLIC to quantify and mitigate these errors. First, we implement an ensemble RPC instead of using an ensemble standard cross-correlation, and develop an SLIC optimal filter that maximizes the correlation strength in order to reliably and accurately detect the correlation peak representing the most probable average displacement of the nanoparticles. Secondly, we developed an analytical model of the SLIC measurement bias error due to image scanning of diffusion-dominated tracer particles. We show that the bias error depends only on the ratio of the mean velocity of the tracer particles to that of the laser scanner and we use this model to correct the induced errors. We validated our technique using synthetic images and experimentally obtained SLIC images of nanoparticle flow through a micro-channel. Our technique reduced the error by up to a factor of ten compared to other SLIC algorithms for the images tested in this study. Moreover, our optimized RPC filter is reducing the number of image pairs required for the convergence of the ensemble correlation by two orders of magnitude compared to the standard cross correlation. This feature has broader implications to ensemble correlation methods and should be further explored in depth in the future.

**Nomenclature**

$X_{f,p}$    Position of the $p^{th}$ particle from a confocal scanned image (frame f)

$X_{f,p}^0$    Instantaneous position of the particle at the beginning of the scan

$X'_{f,p}$    Random movement caused by Brownian motion in discrete time

$t_{f,p}$    Actual time elapsed for a scanner to detect particle before the end of the scanning

$\Delta t$    Time elapsed per frame (line)

$\Delta t_{f,p}$    $t_{f+1,p} - t_{f,p}$

$\Delta X_{f,p}$    $X_{f+1,p} - X_{f,p}$

$\Delta X_{f,p}^0$    $X_{f+1,p}^0 - X_{f,p}^0$

$U_f$    Velocity of the fluid

$U_s$    Velocity of the laser scanner

$D$    Diffusion coefficient $\left[\frac{\mu m^2}{\mu s}\right]$

PIV    Particle image velocimetry

RPC    Robust phase correlation

SLIC    Scanning laser image correlation

SLICR    Scanning laser image correlation - Robust phase correlation

RICS    Raster image correlation spectroscopy

FCS    Fluorescent correlation spectroscopy

CLSM    Confocal laser scanning microscopy

SNR    Signal-to-noise ratio

**Introduction**

Scanning laser image correlation (SLIC) [1] is an image-based technique for measuring the velocity of flow-tracer particles suspended in a liquid using confocal laser scanning microscopy (CLSM). The fundamental operating principle of SLIC is equivalent to that of traditional particle image velocimetry (PIV): a sequence of images is acquired with a known time separation, and the displacement of particle patterns between consecutive images is measured using cross-correlations. However, the specialized imaging arrangement of CLSM introduces recording artifacts that preclude the straightforward application of PIV algorithms to CLSM images. The objective of this work is to identify these specific artifacts, quantify their effects on the accuracy of correlation-based velocity measurements, and subsequently develop and demonstrate algorithms that mitigate them.

PIV is well developed for measuring fluid velocity fields in planar or volumetric regions of interest [2, 3], and its extension to light microscopy (µPIV) has been widely and successfully adopted for interrogating micro-scale flows [4, 5]. However, the diffraction-limited optics of light microscopy limit the spatial resolution of traditional µPIV measurements to approximately 0.1 µm. This precludes the use of µPIV for interrogating nanoscale flow and subcellular kinematic processes such as nanoparticle advection and diffusion across extra-cellular matrices or the transport of fluorescently labeled molecules across cell membranes.

In contrast, confocal microscopes overcome the diffraction barrier by collecting light through a small pinhole placed in line with the objective lens. The pinhole images a single point in the specimen, and rejects out-of-focus light by blocking rays that do not emanate from the plane of focus. Two-dimensional images are formed point-by-point by scanning the pinhole/objective over the region of interest. This arrangement enables spatial resolution on the order of nanometers [1, 6]. On this basis, CLSM is often a preferred technique for imaging small-scale features of fixed cells [6-9].

However, the high spatial resolution of CLSM comes at the expense of temporal resolution. The formation of images by scanning is slow compared to bright field arrangements and introduces "motion blur" when imaging specimens that move with velocity similar to that of the scanner.

CLSM is therefore not widely used to interrogate many of the dynamic transport processes of interest to cellular biology.

Despite this shortcoming, Rossow et al. used PIV-like algorithms to measure the velocities of flow tracer particles from one-dimensional CLSM images (i.e., single line-scans) [1]. In this study, the researchers captured CLSM images of 0.1μm particles flowing in a microfluidic channel, and in vivo in the circulatory systems of transparent embryonic zebra fish. However, assessing the accuracy of their results is difficult because no estimates of error or uncertainty were reported. Moreover, their report does not address the two major sources of error to which correlation-based velocity measurements from CLSM images are subject: a random error due to the Brownian motion of small tracer particles, and a bias error due to the formation of images by progressive scanning.

The positions of tracer particles in successive CLSM images are related by the velocity of the fluid as well as a random displacement due to Brownian motion. As the time between image pairs increases, these random displacements cause the particles' positions to deviate from those of the fluid pathlines. Such a stochastic process introduces a random error into the cross-correlation based velocity estimate. Moreover, because the images of multiple particles contribute to each cross correlation, and because their random displacements are by definition uncorrelated, this effect manifests as randomly distributed peaks in the cross correlation, as illustrated in Figure 1.

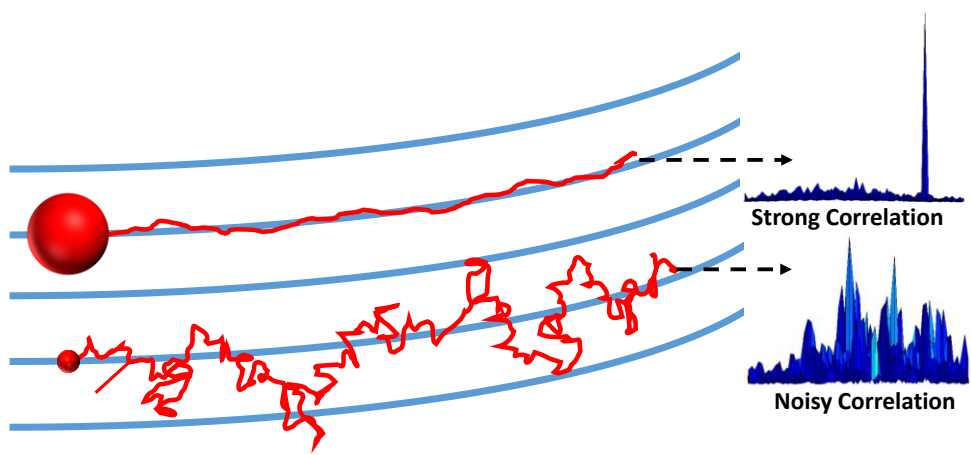

**Figure 1. Schematic of expected contributions of Brownian motion to particle motion in flow, in which correlation quality decreases while imaging small-sized nanoparticles.**

The deleterious effects of diffusion on the cross correlation are not unique to CLSM images, but similarly affect any system that images tracer particles with significant diffusion. Typically, these effects are dealt with by summing the correlations of many pairs of images so that the zero-mean errors of diffusion average out, and the remaining correlation peak is interpreted to represent the best estimate of the time-averaged displacement of the tracer particles [3, 10]. This summing of correlations is known as the "ensemble correlation," which has been shown to reduce random error in micro particle image velocimetry (µPIV) experiments compared to single-pair correlations [2, 5]. The accuracy of the ensemble correlation–based displacement estimate depends on the number of image pairs used, or more appropriately, the total number of pixels contributing to the correlation. Therefore, the number of frames (defined for this study as the number of horizontal single pixel line scans) is critical for accurate velocity measurements using diffusion-dominated particles as flow tracers.

In addition to the effects of diffusion, a second significant error source in SLIC arises from the formation of images by laser scanning. In SLIC, a laser beam scans in one direction across the imaging domain, and light is collected from one point at a time (Figure 2a). Unlike traditional PIV images, a CLSM image does not represent an instantaneous snapshot of the entire flow field. Rather, each pixel is recorded sequentially in time, with a delay between pixels that is usually significant compared to the velocity of the tracer particles. Because of this, the images of the particles within CLSM photographs (and their positions between recordings) are distorted according to the ratio of the particle velocity to the CLSM scanning velocity. As illustrated in Figure 2, this effect manifests as a bias error in the velocity measurement. Specifically, the measured displacement is biased toward zero as the ratio between the particle and scanning velocity approaches unity, while the limiting case of infinitely fast scanning velocity is tantamount to traditional "snapshot" imaging. Figure 2b illustrates this bias error with the example of a one-dimensional line scan (Figure 2a) acquired with a single particle flowing under constant velocity.

A protocol for minimizing both the random and bias errors to which SLIC-based flow velocimetry is subject will enable the development of more robust systems for using confocal microscopy to characterize nanoscale flow kinematics. Currently, however, no such protocol exists. Here, we propose a novel processing method that combines the Robust Phase Correlation (RPC) [11, 12] and ensemble correlation [4, 5] with SLIC [1], and develops and implements a framework for bias error correction to overcome these limitations.

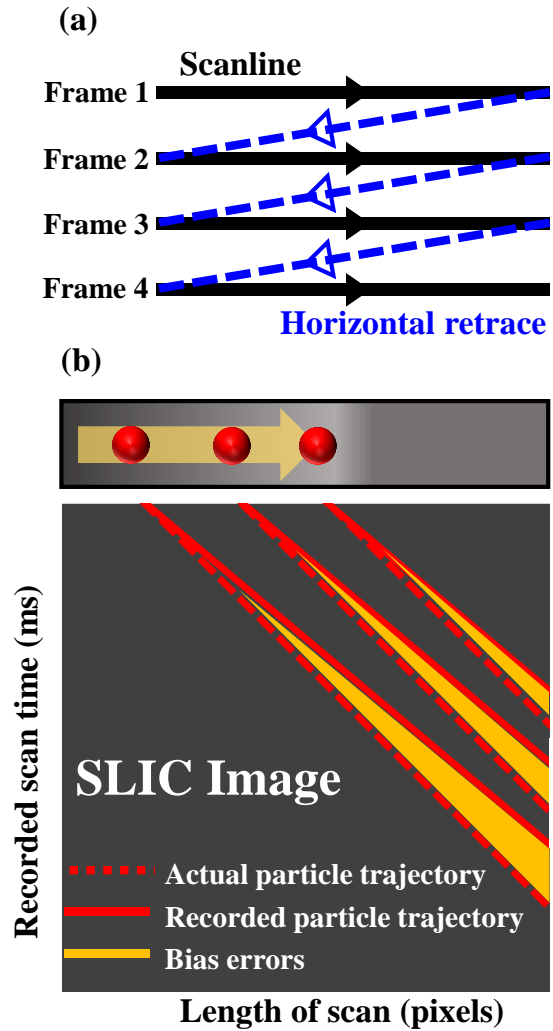

**Figure 2 (a) Illustration of single line scanning (Rossow et al. [1]) and (b) presence of bias errors at a particular ratio of the scanning and fluid velocities. This example illustrates the bias error alone, without the contributions of random error due to diffusion.**

**Robust Phase Correlation (RPC)**

The RPC algorithm extends traditional PIV by applying a filter to the generalized cross correlation (or "phase correlation") in the Fourier domain. The purpose of the "RPC filter" is to increase the signal-to-noise ratio (SNR) of the phase correlation by preferentially weighting the wavenumbers that contribute to the true displacement estimate, and attenuating those that contribute to random errors (e.g., noise due to image digitization). The Gaussian shape of this filter was originally based on a theoretical model of the image formation process in traditional PIV experiments. Eckstein *et al.* showed that the RPC filter reduced random and bias errors in simulated and experimental PIV measurements compared to the standard cross correlation (SCC) and the unfiltered phase correlation (PC) [11-13].

In these studies, the RPC filter improved not only planar PIV, but also µPIV using light microscope images that exhibited comparatively greater background noise and lower SNR than traditional measurements [11, 14, 15]. This suggests that the spectral characteristics of those light microscope images were to some extent congruent with the theoretical model of RPC. However, the image formation process of CLSM is significantly different from that of traditional cameras, and it is therefore likely that their spectral characteristics, too, will differ. Nonetheless, previous results suggest that an RPC-like filter, of yet undetermined width, will likewise improve the correlation SNR and reduce errors in CLSM velocimetry.

**Bias Correction**

The particle image positions in CLSM are influenced by three primary factors: fluid velocity ($U_f$), scanner velocity ($U_s$), and random displacement (X') caused by Brownian motion. Equation (1) represents the positions of 1-D confocal images of a tracer particle in a uniform, unidirectional flow. The $X_{f,p}$ is the position in the image of the $p^{th}$ particle from the recorded confocal line scan ($f^{th}$ frame). $X_{f,p}^0$ refers to the initial position of the $p^{th}$ particle in space at the beginning of the $f^{th}$ frame, and $t_{f,p}$ is the time elapsed between the beginning of the $f^{th}$ frame and the moment the scanner arrives at the position of the $p^{th}$ particle ( i.e., $X_{f,p}/U_s$) . $X'_{f,p}$ represents the random displacement of the $p^{th}$ particle during the interval $t_{f,p}$. Subsequently, the same definition is assigned for the next consecutive position and time variables used for each consecutive frame (f = 1, 2, 3 …).

$$X_{f,p} = X_{f,p}^0 + U_f \cdot t_{f,p} + X'_{f,p} \quad (1)$$

Subsequently, the displacement of particles between an image pair is given by Equation (2):

$$X_{f+1,p} - X_{f,p} = X^0_{f+1,p} - X^0_{f,p} + U_f \cdot (t_{f+1,p} - t_{f,p}) + X'_{f+1,p} - X'_{f,p} \quad (2)$$

The difference $X_{f+1,p} - X_{f,p}$ will be referred to as $\Delta X_{f,p}$ and $\Delta t_{f,p} = t_{f+1,p} - t_{f,p}$, which is the interval between the times at which the scanner reaches the position of the $p^{th}$ particle in subsequent frames ($\frac{\Delta X_{f,p}}{U_s}$). Initial positions $X^0_{f+1,p} - X^0_{f,p}$ (referred as $\Delta X^0_{f,p}$) can be expressed alternatively as Equation (3), $\Delta t$ is the recorded elapsed time per frame from CLSM and $X^{0\prime}_{f,p}$ is the random displacement that occurred between the beginnings of frames f and f+1.

$$\Delta X^0_{f,p} = U_f \cdot \Delta t + X^{0\prime}_{f,p} \quad (3)$$

Substituting Equation (3) to Equation (2), we get

$$\Delta X_{f,p} = U_f \cdot \Delta t + U_f \cdot \frac{\Delta X_{f,p}}{U_s} + X'_{f+1,p} - X'_{f,p} + X^{0\prime}_{f,p} \quad (4)$$

Additionally, the position of the particle in the image can also be expressed in terms of the laser scanning velocity ($U_s$) which are represented on Equation (5) and Equation (6) below.

$$X_{f,p} = U_s \cdot t_{f,p} \quad (5)$$
$$\Delta X_{f,p} = U_s \cdot \Delta t_{f,p} \quad (6)$$

The unknown variables from Equation (4) are $U_f$ and the combined random displacement, $X'_{f+1,p} - X'_{f,p} + X^{0\prime}_{f,p}$ which occurred over three different time instances. In terms of measurements, we can regard the ensemble averaged $<\Delta X_{f,p}>$ as the most probable displacement of particles estimated by averaging cross-correlation of a confocal scanned image pair over total number of particles (q) per each frame with a sufficiently large total number of frames (m), given by Equation (7).

$$<\Delta X_{f,p}> = \frac{1}{m}\sum_{f=1}^{m} \frac{1}{q}\sum_{p=1}^{q} \Delta X_{f,p} \quad (7)$$

Equation (8) below represents the ensemble average of each term.

$$< \Delta X_{f,p} > = U_f \cdot \Delta t + U_f \cdot \frac{< \Delta X_{f,p} >}{U_s} + < X'_{f+1,p} - X'_{f,p} + X^{0'}_{f,p} > \qquad (8)$$

The ensemble average of the random displacements $< X'_{f+1,p} - X'_{f,p} + X^{0'}_{f,p} >$ will yield zero, due to Brownian motion which can be modeled as normally distributed variable with zero mean [16]. Subsequently, we get Equation (9) with one unknown variable $U_f$

$$< \Delta X_{f,p} > = U_f \cdot \Delta t + U_f \cdot \frac{< \Delta X_{f,p} >}{U_s} \qquad (9)$$

After rearranging, the fluid velocity $U_f$ can now be decoupled from the measured ensemble averaged displacement $< \Delta X_{f,p} >$, which consists of the bias error due to the effect of scanning, given by Equation (10)

$$U_f = \frac{< \Delta X_{f,p} >}{(\Delta t + < \Delta X_{f,p} >/U_s)} \qquad (10)$$

**Processing Algorithm**

Equation (10) implies that the underlying fluid velocity can be accurately estimated when measurements of the tracer particles' displacements are statistically converged and no longer subject to significant random fluctuations. Hence, it is necessary that the statistical image correlation converge within a finite number of ensembles. In order to achieve that we develop a processing algorithm (Figure 3), referred to thereon as SLICR (SLIC with RPC), which differs from conventional ensemble SCC and RPC in terms of two specific steps. First, while the original RPC filter was based on a theoretical model of digital image formation, we instead apply a modified Gaussian-shaped spectral filter whose width (standard deviation) was chosen to be optimal for minimizing error and accelerating convergence in ensemble correlations of computer-generated CLSM images. Secondly, we use the previously described analytical model of CLSM image formation to account for the measurement bias error due to the scanning process. A complete description of our SLICR algorithm follows below.

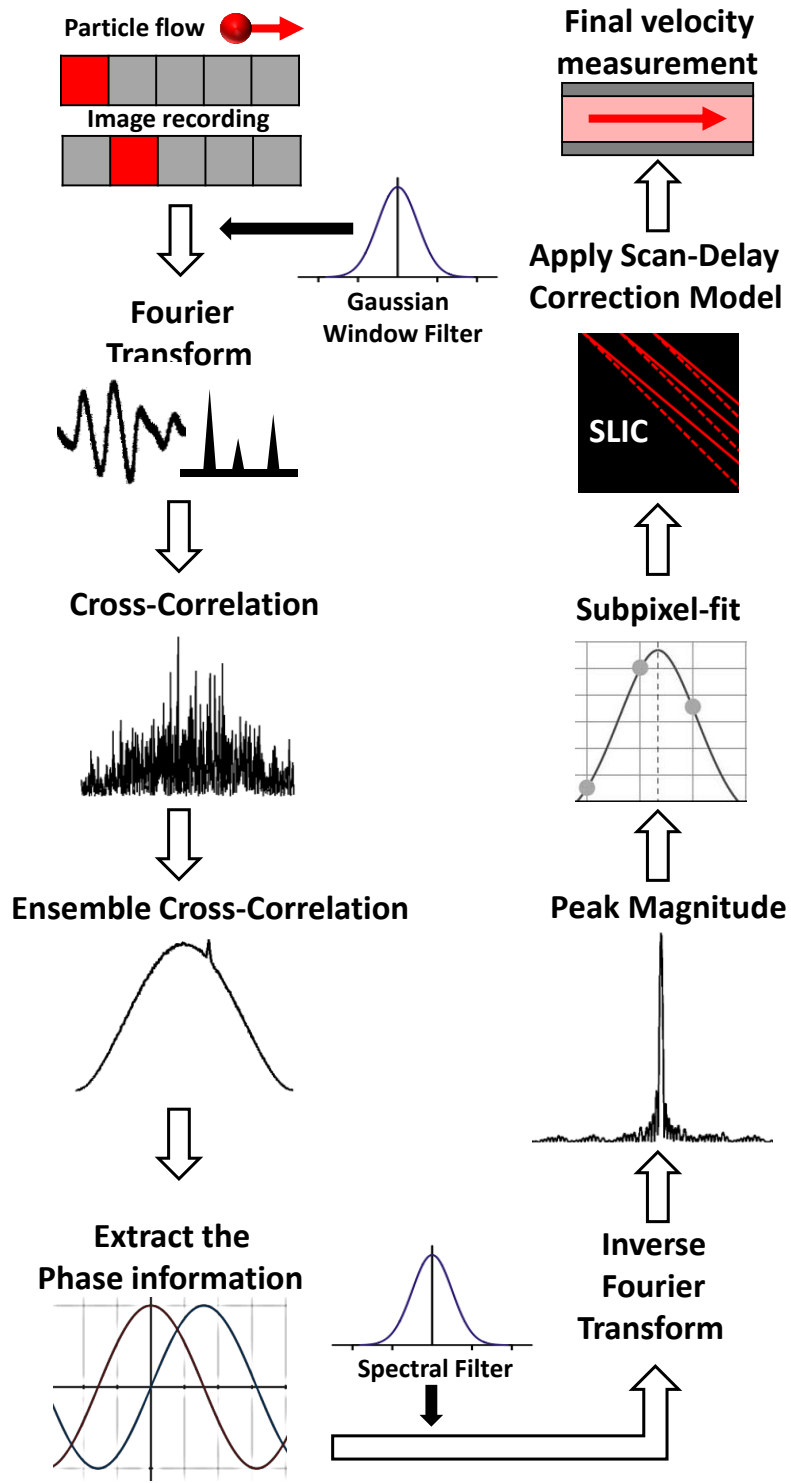

**Figure 3: Our SLICR processing algorithm for measuring time-averaged velocity of particles imaged by confocal microscopy**

The details of the first five steps of our processing algorithm involving a Gaussian apodization filter, Fourier transforms (FTs) and ensemble correlation are discussed in detail by Eckstein et al (2008), which yields the raw (unfiltered) phase correlation [12]. The raw phase correlation is then multiplied by a Gaussian-shaped weighting function (the "RPC" filter) whose value is unity at the zero-wavenumber pixel and exponentially decays at larger wave numbers. The only parameter for this filter is its standard deviation (or "width"), and the method by which we define the standard deviation is described later. We refer to this filtered phase correlation as the "robust phase correlation." After filtering the phase correlation, we calculate its inverse FT, which yields the characteristic "peak" whose position indicates the most probable displacement of the image pattern (subject to the previously described bias error). We estimate the sub-pixel location of the correlation peak as the maximum value of the continuous Gaussian function that best fits the points surrounding it. Finally, we apply the previously described bias correction to this sub-pixel displacement, which yields our best estimate of the time-averaged displacement of the particles that were photographed.

**Determination of the RPC Filter Width**

In the absence of an analytical model of the noise characteristics of CLSM imaging, we optimized the width of the Gaussian RPC filter using Monte-Carlo error analysis of computer-generated CLSM line scans ("images") of the steady unidirectional flow of tracer particles (the image generation procedure is described subsequently). We randomly varied the ratio of the particle velocity to scanner velocity (the "velocity ratio") from 0.01 to 0.1, and varied the rate of diffusion from 10 to 60 pixels$^2$/frame. Each velocity measurement was derived from of an ensemble correlation of $2\times10^5$ pixel counts, which we determined was adequate for all measurements to converge. After calculating the raw (unfiltered) phase correlation of each ensemble, we parametrically varied the standard deviation of the Gaussian RPC filter from zero (essentially a Dirac delta function) to the width of the correlation (i.e., nearly flat). We assessed the measurement error for each filter width as the absolute difference between the ground truth particle displacement and the SLICR-measured displacement after bias correction. As shown in Figure 4, our analysis indicates that error was minimized for a filter diameter of about 33 pixels (equivalent to the filter standard deviation of about 18 pixels).

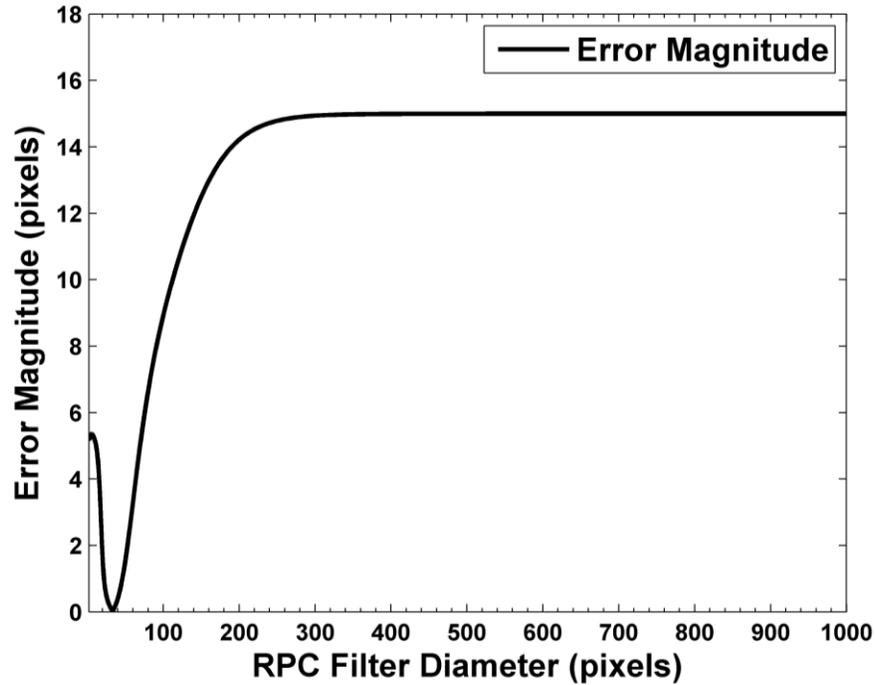

**Figure 4: Relationship between RPC filter diameter and displacement error magnitude, as determined by Monte-Carlo analysis of synthetic computer generated confocal microscope images of tracer particles under steady unidirectional flow.**

**Assessment of SLICR Algorithm Performance: Synthetic Images**

**Synthetic Image Generation**

CLSM images of particles under unidirectional flow were rendered by sampling synthetic particle flow fields pixel-by-pixel, scanning horizontally, while continuously advecting the underlying tracer particles. The interrogated flow field was simulated over a 2D domain, and was seeded with tracer particles with random initial vertical and horizontal coordinates. The domain was discretized into 1,000 horizontal interrogation regions (or "pixels"), each of which represented a single station at which the CLSM scanner sampled the field. Images were formed progressively by sampling the flow field at each pixel, whose resulting brightness was determined by integrating the contributions of nearby Gaussian-shaped particles according to the theory of Adrian and Olsen [17]. The positions of the particles were updated between integrations according to a prescribed uniform horizontal velocity (in the direction of scanning) as the scanner progressed. In our simulations, the fluid velocity was varied from 0 to 100 pixels per line (by adjusting the fluid and scanner velocity). The time elapsed between each pixel integration corresponded to the input scanner velocity divided by the size of the domain. Additionally, diffusion was modeled as pseudo-random displacements in the horizontal and vertical directions. This pseudo-random displacement was generated from a normal distribution with a mean of zero and a standard deviation of $\sigma = \pm\sqrt{2 \cdot D \cdot \tau}$, where $D$ is the Stokes-Einstein diffusion coefficient [18] in $\mu m^2/\mu s$ and $\tau$ is in the elapsed time per frame in seconds. We chose a diffusion coefficient of 70mm$^2$/s and 4.9 mm$^2$/s to modeling 7 nm and 100nm particles, respectively. The particles were modeled to be suspended in water at 25°C, which allowed the particles to move freely due to diffusion and advection. In our simulations, a single horizontal line was scanned repeatedly to match the behavior of the experimental apparatus described later. Uncorrelated Gaussian noise was added to each pixel to represent the effects of imaging noise, with a mean value of 0 and standard deviation of 10% of the pixels' saturation intensity. To maintain consistency between our simulations and experiments, we simulated particle diameters of 7nm and 100nm; these corresponded to particle image diameters of 2 and 6 pixels, respectively. These particle image diameters ($D_p = \sqrt{2} \cdot D_A$) were calculated by using the autocorrelation diameter ($D_A$) measured from the experimental images obtained with nanoparticles [10].

**Experimental Assessment of SLICR Algorithm Performance: Micro-channel Flow.**

To evaluate the performance of our SLICR algorithm on real data, we collected confocal images of nanometer-sized tracer particles suspended in water flowing through a plastic microfluidic channel of rectangular cross section (μ–Slides I Luer, ibidi Inc). Figure 5 illustrates the overall experimental system and imaging location. The dimensions of the channel were 0.1mm (depth) x 5mm (width) x 50mm (length). Polystyrene microspheres (0.1 μm diameter; Fisher Scientific) and CdSe/ZnS quantum dots (7 nm diameter; Sigma-Aldrich, 694614) were used as tracer particles. The channel was filled with a suspension of particles in water with a seeding density of 1mg/100mL. The volumetric flow rate through the channel was controlled by a syringe pump (Harvard Apparatus), and ranged from 0.005 to 0.5 μL/s. The interrogation region was located near the center of the channel (x and z-axis), with 11 different positions (44, 40, 30, 20, 10, 0, -10,

-20, -30, 40 and -44μm) spaced along y-axis (Figure 5). The nominal expected flow velocities at the interrogation spot ranged from 10 to 1000 μm/s.

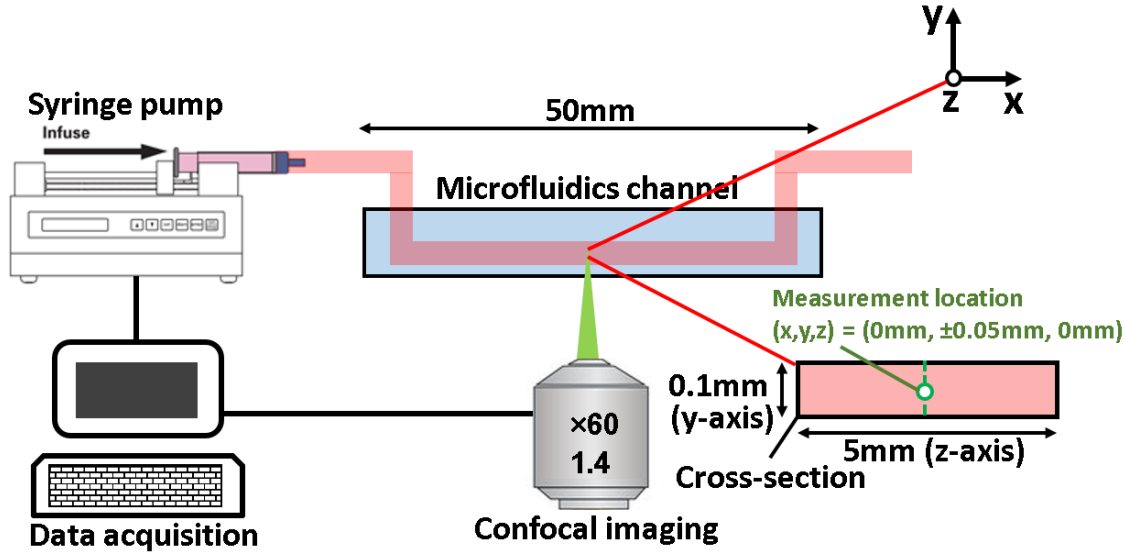

**Figure 5: Schematic of the experimental setup used to obtain confocal microscope images of nanometer-sized tracer particles flowing in water through a microchannel.**

A Nikon A1R scanning laser confocal microscope (Nikon Corporation, Tokyo, Japan) was used to photograph the flow through the microfluidic channel. The channel was viewed through a 60x objective lens (numerical aperture NA = 1.4, working distance of 0.2 mm), and illuminated by an argon ion laser (561 nm wavelength). The scanned path was a line 512 pixels long, oriented approximately parallel to (and in the same direction as) the mean flow. The dwell time at each pixel was selected between 2.2 – 12.1 μs, with an image magnification of 0.05 μm per pixel. Each trial consisted of 10,000 consecutive scans along the same path. The scan time for each line ranged from 0.1 to 7.7 ms, which includes additional time pausing at the beginning and end of each line. The spatial resolution of the automated traverse (Z step size) was 0.13 μm.

**Quantification of Error**

For both synthetic and experimental images, we assessed the performance of our algorithm using two metrics. First, we quantified the number of line scans required for convergence of the displacement estimate using our SLICR compared to the standard cross correlation (SCC) used in SLIC. Our criterion for convergence was set that the velocities across successive ensemble converge within 0.1 pixels (for 1,000 consecutive estimates) to ensure velocity curve with respect to increasing number of ensemble reach a plateau. The upper bound limit of 0.1 pixels was referenced from the standard deviation of 1,000 displacement estimate (instantaneously cross-correlated) from synthetic CLSM images generated with no effect of diffusion. For the SLICR algorithm, we used the previously optimized RPC filter diameter of 33 pixels. This metric depended only on the behavior of the correlations themselves, and therefore isolated the effects of the RPC filter from those of the bias correction model.

Secondly, we compared the accuracy of converged SLICR-calculated particle velocity estimates with and without application of our bias correction model. For these tests, we parametrically varied the ratio of fluid velocity to scanning velocity from 0.002 to 0.11 (for synthetic data) and 0.01 to about 0.11 (for experimental data). Our metric of accuracy was the absolute difference between the ground truth velocity of particles and the SLICR-measured velocity. For synthetic data, the ground-truth velocity was taken as that prescribed in the simulations. For experimental data, we estimated the ground-truth velocity analytically using the equation for fully developed plane Poiseuille flow, evaluated at the measurement locations that we interrogated.

As an additional point of comparison, we used SCC and SLICR to measure the velocity profile of the channel flow by performing measurements at eleven locations evenly spaced in the depth direction between ±50 μm of the channel centerline. Error was quantified as the difference between the measured and theoretical velocity at each position.

For all converged measurements using experimental images, we estimated the uncertainty of the calculated error by propagating (via the Taylor series expansion method) the known elemental sources of error in our experiment, and the RMS of the measured velocity, through the error equation. The elemental sources of error we considered were the volumetric flow rate delivered by the syringe pump, the physical location of the interrogation region, and the dimensions of the microfluidic channel, whose values were used to calculate the Poiseuille flow velocity profile. The variation of the fluid viscosity was not included in the uncertainty calculation due to the absence of viscosity measurements of the distilled water, which was estimated to be marginal relative to other manufacturing tolerances.

## Results

**Convergence of Measurements: Synthetic Images**

Figure 6 shows a representative comparison of convergence of ensemble SLICR and SCC velocity estimates, for the measured velocity normalized by the expected velocity (corresponding to the measurement with the fixed input velocity of 1000μm/s with dwell time of 2.0μs). Whiskers indicate the 95% confidence interval about the mean velocity ratio (40 data points are shown in the plot in order to distinguish markers and whiskers for each case across large range of ensemble pixel counts on log-scale). The expected velocity (blue) was estimated by using Equation (10) with known input fluid velocity and scanner velocity.

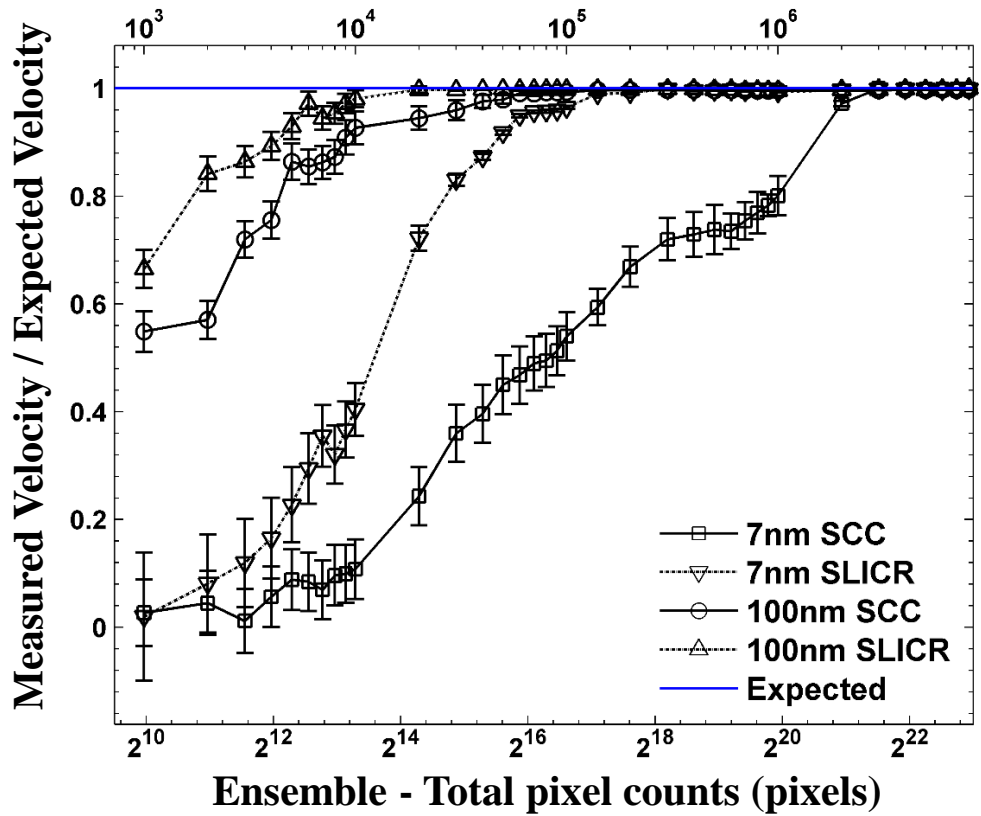

**Figure 6: Convergence behavior of SLICR and SCC algorithms for synthetic images of 7nm and 100nm flow tracer particles for the velocity measurement normalized by the expected value, corresponding to the measurement with the input velocity of 1000μm/s with dwell time of 2.0μs**

**Table 1: Convergence estimate for the measurement with synthetic images**

|  | Convergence (total pixel counts) | |
|---|---|---|
| **Particle size** | *SCC* | *SLICR* |
| *7nm* | $3.0 \times 10^6$ | $1.4 \times 10^5$ |
| *100nm* | $6.0 \times 10^4$ | $2.0 \times 10^4$ |

The results show that the SLICR correlation converges after an ensemble length of about $2.0 \times 10^4$ pixel counts (or 20 line scans of 1000 pixels in our simulations) for 100nm particles, and about $1.4 \times 10^5$ pixel counts (or 140 line scans) for 7nm particles (Table 1). The difference in convergence behavior between particle sizes is due to the increased contribution of diffusion to the displacements of the smaller particles. Meanwhile, convergence of the SCC algorithm required about $6.0 \times 10^4$ and $3.0 \times 10^6$ pixel counts for 100nm and 7nm particles, respectively. This difference suggests that the filtered phase correlation of SLICR increases the robustness of the measurements against the deleterious effects of using flow tracers that exhibit significant Brownian motion.

The general trends for all cases show that the velocity estimates increase with increasing ensemble lengths until the plateau region is reached. For SCC with 7nm particles, the mean velocity ratio values were close to zero (with large 95% confidence intervals indicating the frequent change in the peak detection) over smaller number of ensemble up to $2^{14}$ (16384) pixel counts. Such behavior represents the significant presence of random errors in the measurement. Subsequently, the deviation in the mean measurement for all cases decreased towards larger number of pixel counts being averaged.

These convergence estimates are most significantly influenced by the diffusion coefficient in the system as described previously, but also depend (to a lesser degree) on the ratio of fluid velocity to scanner velocity.

Figure 7 shows representative convergence behaviors of SLICR with respect to different fluid to scanner velocity ratios. The results show that the SLICR correlation with 100nm particles with the velocity ratio of 0.01, 0.02 and 0.03 converged after an ensemble length of about $0.9 \times 10^4$, $1.0 \times 10^4$ and $1.0 \times 10^4$ pixel counts, respectively. For 7nm particles, SLICR correlation with the velocity ratio of 0.01, 0.02 and 0.03 converged after an ensemble length of about $12\times10^4$, $14\times10^4$ and $12\times10^4$ pixel counts, respectively.

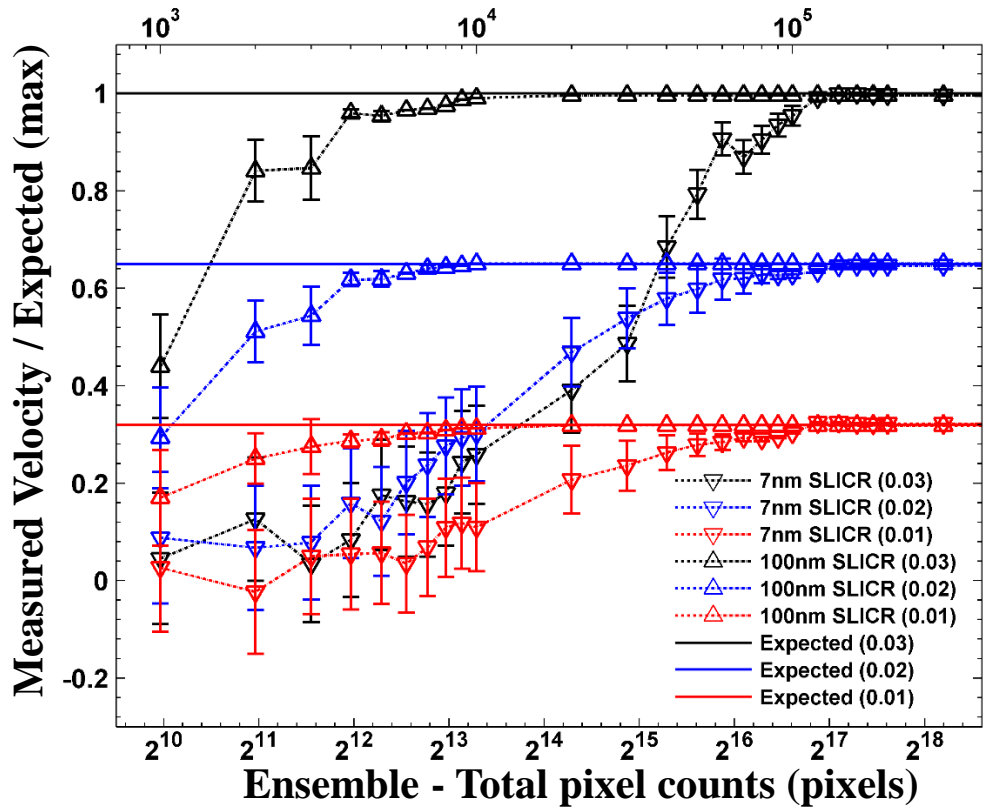

**Figure 7: Convergence behavior of SLICR algorithms for synthetic images of 7nm and 100nm flow tracer particles for the velocity measurement normalized by the expected value (maximum). The fluid to scanner velocity ratio are indicated for each legend.**

**Bias Correction: Synthetic images**

Figure 8 shows the accuracy improvement of our bias correction model on the individually converged ensemble SLIRC velocity estimates in synthetic images. The mean numbers of pixels required for convergence were $1.4\times10^4$ and $1.3\times10^5$ for 100nm and 7nm particles, respectively, for the range of different fluid to scanner velocity ratios shown on Figure 8. As predicted by Equation (10), the bias error increases with the ratio of the fluid velocity to that of the scanner (i.e., relatively slow scanning), with a maximum error magnitude of about 11 pixels. Conversely, high scanning velocities decreased the bias error, approaching the behavior of traditional "snapshot" imaging. The filled markers in Figure 8 show the remaining error after application of our bias correction model.

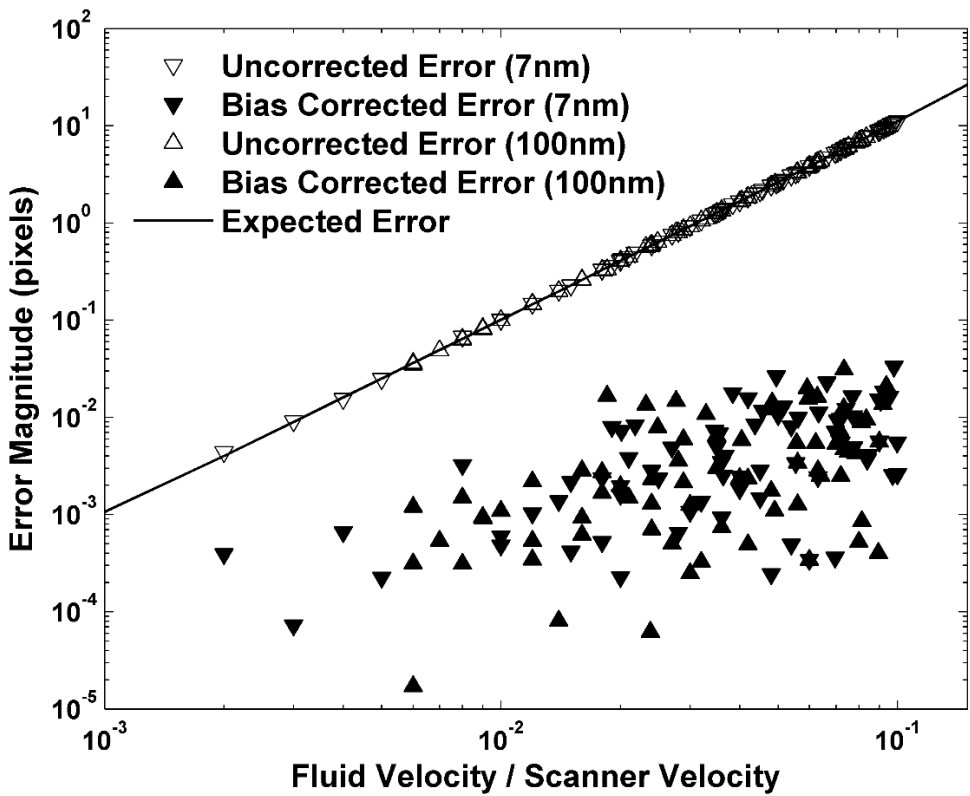

**Figure 8: Absolute velocity errors with respect to the velocity ratio (fluid/scanning) measured from the individually converged ensemble SLICR measurements.**

These results indicate that applying our bias correction reduced the measurement error to within $3.3 \times 10^{-2}$ pixels for synthetic data. The 95% confidence interval about the mean error was within 0.01 pixels. The remaining scatter in the errors of the corrected measurements is likely due to the various unaccounted random sources of error to which PIV-like algorithms are subject (the subpixel fitting algorithm, etc.), the mitigation of which are beyond the scope of this study.

**Experimental Demonstration**

Figure 9 shows representative CLSM images of 7nm and 100nm particles in water flowing through our microfluidic channel. These images are qualitatively similar to those presented previously by Rossow *et al*. [1], and exhibit the familiar diagonal patterns that characterize the line-scan imaging of moving particles. Note that these are not two-dimensional images: instead, each row of pixels represents a single line-scan (starting from the top), and subsequent line scans across the same physical domain appear as consecutive rows in the images. In other words the slopes of the 100nm particles' trajectories shown in Figure 9(a) represent their velocities, subject to the bias error discussed previously.

The images of 7nm particles shown in Figure 9(b) lack any clearly discernable patterns or features, and they appear more like random noise. This is due to the significantly increased Brownian motion and dimmer images of the smaller nanoparticles. This image is demonstrative of the challenge inherent in preforming PIV measurements of such small and diffusion-dominated flow tracers.

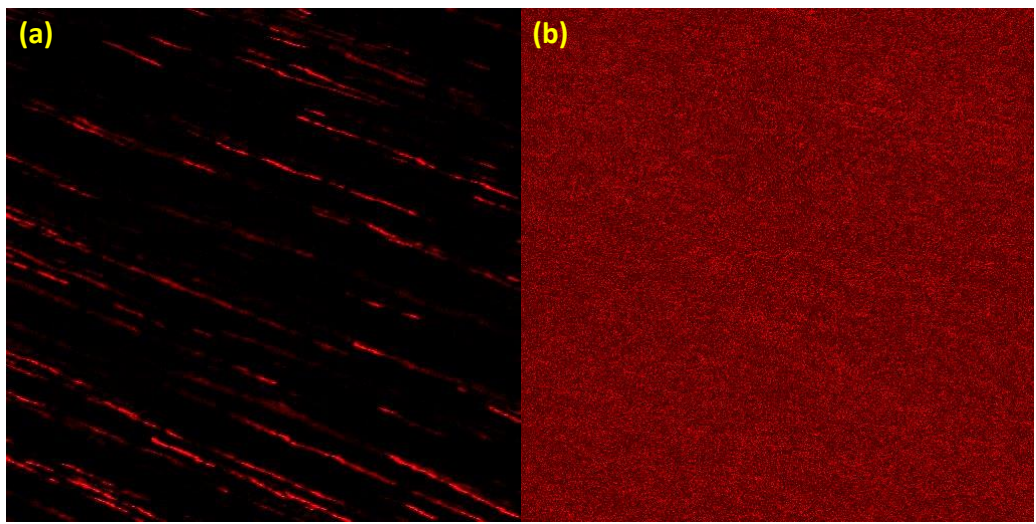

**Figure 9: Line scanning laser confocal microscope images of (a) 100nm particles and (b) 7nm particles suspended in water, subject to flow aligned with the direction of scanning.**

**Convergence Estimation: Experimental Images**

Figure 10 illustrates the convergence behavior of ensemble correlations for experimentally obtained images. The measured velocity is normalized by the expected velocity (corresponding to the measurement with the fixed input velocity of 1000μm/s with dwell time of 2.0μs). As with the synthetic data, the SLICR algorithm significantly reduced the number of pixel counts (or line scans) required for convergence of the correlations for both 7nm and 100nm particles compared to the standard correlation (Table 2). In this case, the ensemble SLICR converged after about $1.0\times10^4$ and $1.3\times10^5$ pixels for 100 nm and 7 nm particles, respectively (in contrast to $2.0\times10^4$ and $1.4\times10^5$ pixels for synthetic images). Meanwhile, the standard ensemble correlation required larger number of pixels to converge for both particle sizes – $5.0\times10^4$ pixels for 100 nm particles, and $4.0\times10^6$ for 7 nm particles, compared to $6.0\times10^4$ and $3.0\times10^6$ pixels for synthetic images.

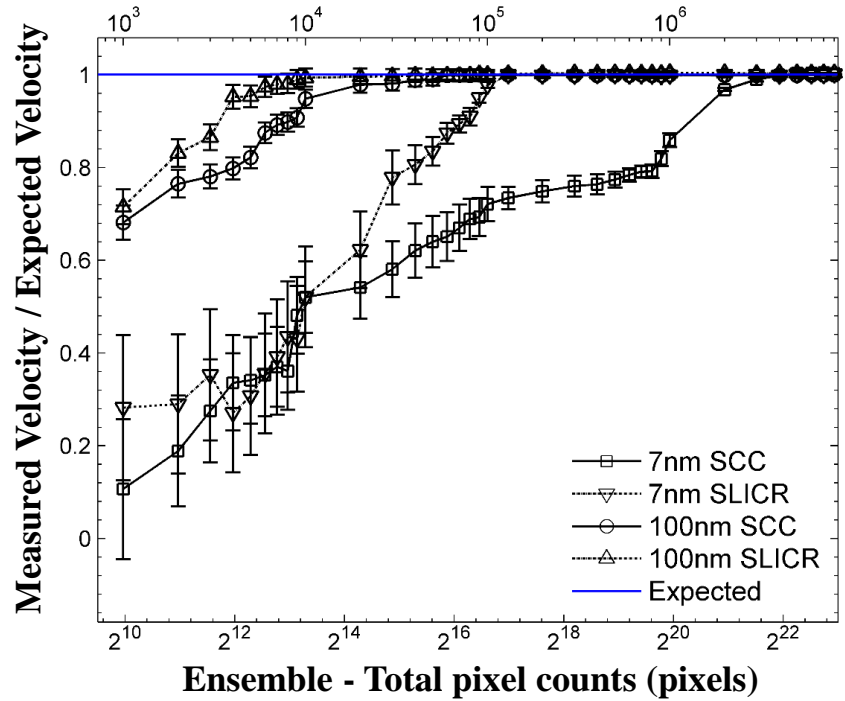

**Figure 10: Convergence behavior of SLICR and SCC algorithms for experimental images of 7nm and 100nm flow tracer particles for the velocity measurement normalized by the expected value, corresponding to the measurement with the input velocity of 1000μm/s with dwell time of 2.0μs**

**Table 2: Convergence estimate for the measurement with experimental images**

|  | **Convergence (total pixel counts)** | |
|---|---|---|
| **Particle size** | *SCC* | *SLICR* |
| *7nm* | $4.0\times10^6$ | $1.3\times10^5$ |
| *100nm* | $5.0\times10^4$ | $1.0\times10^4$ |

Figure 11 illustrates the convergence behavior of SLICR with respect to different fluid to scanner velocity ratios for the experimental data. The results show that the SLICR correlation with 100nm particles and velocity ratios of 0.01, 0.02 and 0.03 converged after an ensemble length of about $1.0\times10^4$, $1.0\times10^4$ and $2.0\times10^4$ pixel counts, respectively. For 7nm particles, SLICR correlation with the velocity ratio of 0.01, 0.02 and 0.03 converged after an ensemble length of about $1.4\times10^5$, $1.4\times10^5$ and $1.6\times10^5$ pixel counts, respectively.

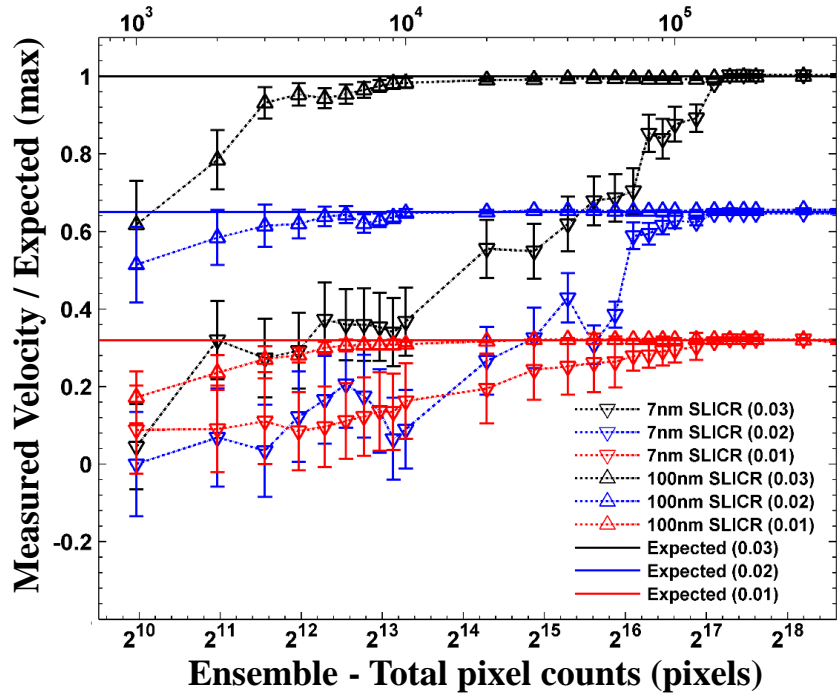

**Figure 11: Convergence behavior of SLICR algorithms for experimental images of 7nm and 100nm flow tracer particles for the velocity measurement normalized by the expected value (maximum). The fluid to scanner velocity ratio are indicated for each legend.**

Figure 12 compares the theoretical and measured velocity profiles within the micro channel using individually converged SLICR and un-converged SCC with both particle sizes. In these trials, the mean number of pixels required for convergence using our algorithm was $1.6\times10^4$ for the 100nm particles and $1.4\times10^5$ for the 7nm particles. For consistent comparison between methods, the SCC measurements were ensemble-averaged using the same number of pixel counts required to converge the SLICR measurements. As a result, the SCC measurements did not themselves converge. After bias correction, nearly all of the measured velocities fell within the 95% confidence interval about the nominal theoretical velocity profile. These results illustrate the importance of bias correction in CLSM velocimetry measurements. In this experiment, the uncorrected bias error was highest near the centerline of the channel, where the flow velocity is greatest. This is due to the previously discussed relationship between the bias error and the fluid-to-scanning velocity ratio, wherein higher flow velocities result in greater bias error for a fixed scanning velocity. In this case, the uncorrected bias error at the centerline reached about 6 pixels (10% of the nominal velocity) for the converged SLICR measurements, which was reduced to 0.4 pixels with the correction model being applied. Meanwhile, the absolute errors reached about 30 pixels with un-converged SCC measurements for both particle sizes, due to the combined and uncorrected effects of both random and bias errors.

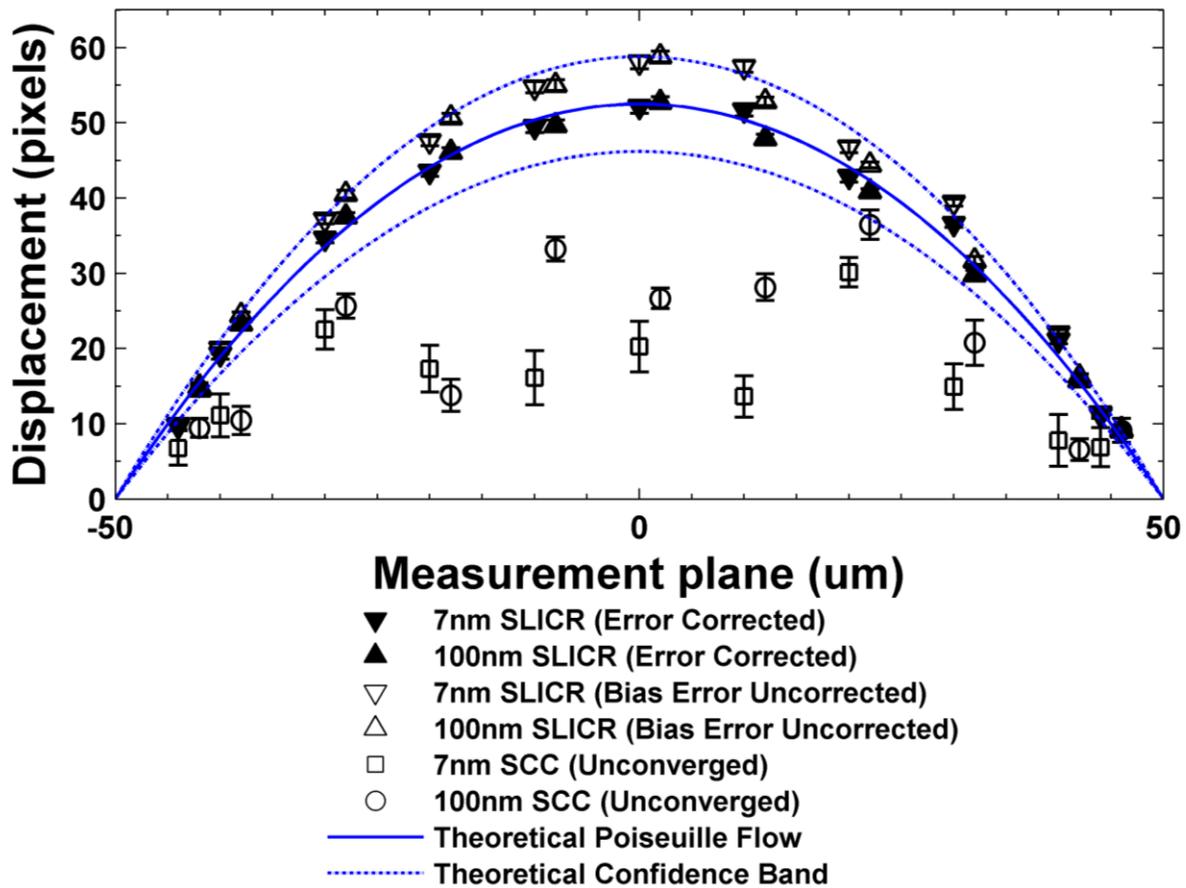

**Figure 12: Comparison of different cases of SLICR and SCC measurements of velocity across the depth of the channel, compared to the theoretical solution for plane Poiseuille flow (flow rate 0.5 µL/sec, fluid-to-scanner velocity ratio between 0.01 and 0.11)**

**Discussion**

Our analysis demonstrates the significant impact of three factors on the accuracy of CLSM based flow velocimetry; namely, the diffusion of the tracer particles, the laser scanning speed, and the velocity of the flow. Subsequently, a processing scheme was developed based on how each factor contributes to the error.

The Brownian motion of tracer particles is the primary driver of random errors in these correlation-based measurements of velocity. This effect manifests as a broad Gaussian-shaped correlation in the ensemble-averaged standard correlation. As other researchers have pointed out, the reason for this shape is that the cross correlation of particle images represents a probability density function of the different particle displacements that contributed to the measurement [5, 10, 19, 20]. In the case of Brownian motion, the PDF of displacements is itself a zero-mean Gaussian function. If the velocities of tracer particles are dominated by diffusion on the time scale of the measurement, then so does this broad Gaussian shape dominate the shape of the cross correlation, as shown in Figure 13a.

This degrades the ability of peak-searching algorithms to correctly identify the comparatively small correlation peak corresponding to the mean background velocity of the flow. The ensemble correlation mitigates this effect because the "true" correlation peak grows to prominence after a sufficient amount of information (e.g., pixels or images) has contributed to the measurement. In contrast, the RPC approach recognizes that the random velocities due to diffusion and the mean background velocity are carried by different wave numbers in the Fourier domain of the phase correlation. More specifically, the RPC filter inherently assumes that the true mean displacement is carried by the lower wave numbers in the phase correlation, and the displacements due to diffusion are carried by the higher wave numbers. Our results show that preferentially weighting the contribution of the lower wave numbers to the phase correlation suppresses the appearance of the broad Gaussian-shape that characterizes diffusion-dominated correlations, while preserving the peak that indicates the true mean velocity of the particle patterns (Figure 13b). Amplifying the relative prominence of this "true" peak accelerates the rate at which the measurement converges with respect to the amount of information contributing to it (in this case, pixels). In this way, our method represents a mechanism to increase not only the accuracy of CLSM velocimetry measurements, but also the temporal resolution by reducing the time needed for image acquisition by more than one order of magnitude compared to existing methods. Moreover, our observations support the RPC filter's fundamental assumption about the spectral anatomy of the phase correlation.

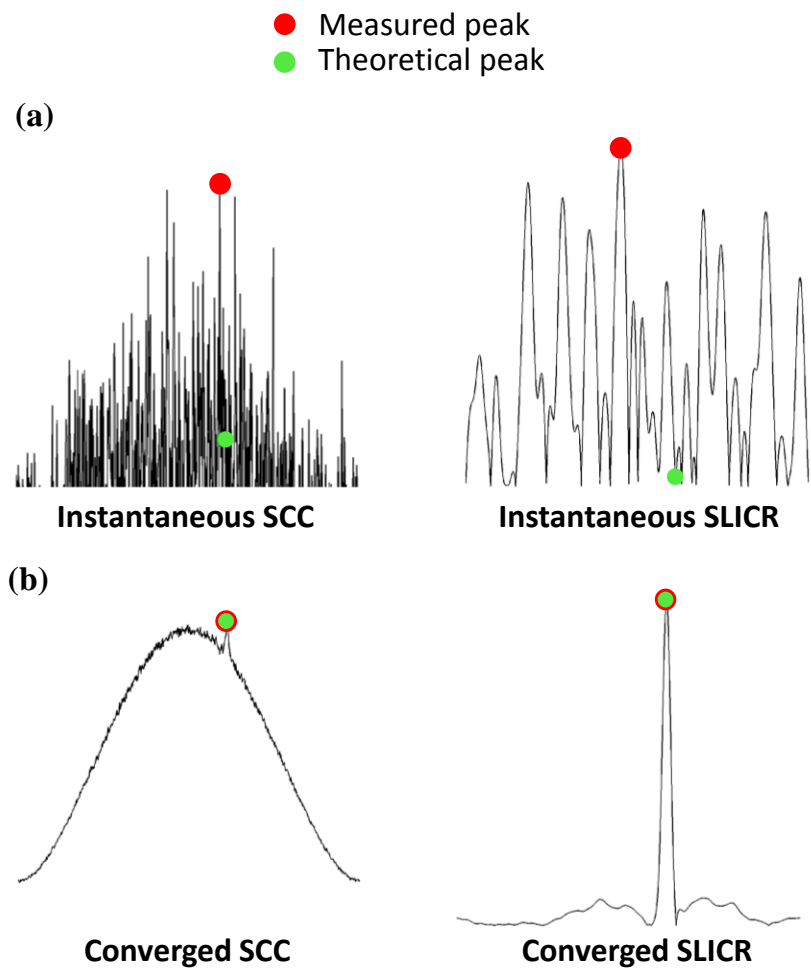

**Figure 13: Representative correlation shape with peaks indicated for (a) the instantaneous SCC and SLICR showing high variability between measured and theoretical peaks (diffusion-dominated signal) and (b) converged SCC and SLICR showing accurately matched measured and theoretical peaks (suppressed diffusion with the SLICR).**

The degree to which Brownian motion affects correlation-based velocity measurements depends not only on the diffusion coefficient itself, but also on the rate of image acquisition (i.e., the microscope's scanning velocity) and the velocity of the underlying flow. Because of the finite scanning velocity of CLSM, the motion of the particles is often significant on the time scale of the acquisition of a single line (or "frame"), which introduces an imaging artifact similar to motion blur in traditional cameras. The limit of infinitely fast scanning represents the traditional "snapshot" photography, which is ideally absent of motion blur. In this situation, diffusion affects the relationship of particle positions between scans, but not within individual scans. The severity of the decorrelating effect of diffusion on the inter-frame positions of the particles depends on the diffusion coefficient and the amount of time separating the two frames (the "inter-frame time" in traditional PIV). Therefore, from the standpoint of minimizing the decorrelating effects of diffusion, it is advantageous to use the fastest possible scanning velocity for this type of CLSM imaging (i.e., the smallest possible inter-frame time). However, as previous researchers have pointed out, the error of PIV measurements relative to the measured velocity (the "relative error") increases when the inter-frame time is so short that the particle displacements are small between frames compared to the fixed sources of error in PIV measurements (discretization, sub-pixel fitting, etc.). Therefore, from the standpoint of reducing the relative error of PIV measurements, it is advantageous to select the largest possible inter-frame time that does not result in unacceptable loss of correlation due to particles leaving the interrogation region [21]. These competing requirements of small inter-frame time to minimize the effects of diffusion and large inter-frame time to minimize the relative error of the measurements represent one of the principle challenges in the design of μPIV experiments. The process of image formation by scanning using confocal microscopes and the use of diffusion-dominated nanoparticles as flow tracers further exacerbate these difficulties by allowing diffusion effects within individual frames (scans) and introducing bias error in the displacement estimate. Our results address these challenges by demonstrating experimentally and through simulation that the effects of Brownian motion are the primary driver of random errors in CLSM-based measurements of particle velocity, and by providing an analytical method by which to mitigate bias errors. Additionally, our analysis verifies the conjecture that these errors should depend on the ratio of the scanning velocity to the flow velocity, rather than on either parameter alone. These insights provide researchers with guidance in the design of similar experiments.

The primary limitation in the present study was the use of a one-dimensional scanning instrument to interrogate flows. We addressed this by constraining our analysis to flows that were themselves one-dimensional and aligned with the scanning axis of the microscope, although in practice the most interesting flows will undoubtedly exhibit 2-D and 3-D structures. We expect that our analysis and principles applied herein can be extended to higher-dimensional measurements and to interrogate 2D flow structures with CLSM, and this work is the subject of our continuing efforts. Moreover, our future analysis will include further investigation to identify additional sources of error in CLSM velocimetry that were not faithfully represented by our simulations, which likely contributed to the discrepancy between the errors we reported between our simulated and experimental results.

Despite these limitations, to our knowledge this work presents the first successful attempt to quantify the error of CLSM-based flow velocimetry and demonstrate improved robustness and accuracy of the method using diffusion-dominated nanoparticles as flow tracers. As a result, we present a theory for and establish a methodology to mitigate the major sources of error and yield reliable velocity measurements with these instruments. More broadly, our research represents a step toward leveraging the exceptional resolving power of confocal microscopes to accurately study the kinematics of nanometer-sized molecules and particles that are of great interest to a wide range of biological systems and cellular mechanics, but have heretofore been obscured by limitations in measurement technology.

## Acknowledgments

The authors would like to acknowledge Dr. Aaron Taylor from the Purdue Bioscience Imaging Facility and Dr. Fred Pavalko for their valuable inputs. This work was supported by grants from the National Institute of Health (1R21EB019646-01 and 1R21AR065659-01), National Science Foundation (106750-PHY-1205642 and 106745-CBET-1335957) and Discovery Park Research Fellowship.